\documentclass[aps,pra,floatfix,twocolumn,showpacs,10pt,longbibliography]{revtex4-2}
\usepackage{amsmath}
\usepackage{url}
\usepackage[ruled,vlined]{algorithm2e}
\usepackage{algorithmic}
\usepackage{graphicx}
\usepackage{dcolumn}
\usepackage{bm}
\usepackage{amssymb}
\usepackage{rotating}
\usepackage[abs]{overpic}
\usepackage{xcolor}
\usepackage{tabularx}
\usepackage{floatrow}
\usepackage{subfig} 
\usepackage[justification=RaggedRight]{caption}
\usepackage{hyperref}
\hypersetup{colorlinks = true, citebordercolor={blue}, linkcolor={blue}, citecolor={blue}, urlcolor={blue}}
\begin{document}











\title{Simultaneous Fermion and Exciton Condensations from a Model Hamiltonian}

\author{LeeAnn M. Sager and David A. Mazziotti}

\email{damazz@uchicago.edu}

\affiliation{Department of Chemistry and The James Franck Institute, The University of Chicago, Chicago, IL 60637}%

\date{Submitted October 26, 2021}


\begin{abstract}

Fermion-exciton condensation in which both fermion-pair (i.e. superconductivity) and exciton condensations occur simultaneously in a single coherent quantum state has recently been conjectured to exist.  Here, we capture the fermion-exciton condensation through a model Hamiltonian that can recreate the physics of this new class of highly-correlated condensation phenomena.  We  demonstrate that the Hamiltonian generates the large-eigenvalue signatures of fermion-pair and exciton condensations for a series of states with increasing particle numbers.  The results confirm that the dual-condensate wave function arises from the entanglement of fermion-pair and exciton wave functions, which we previously predicted in the thermodynamic limit. This model Hamiltonian---generalizing well-known model Hamiltonians for either superconductivity or exciton condensation---can explore a wide variety of condensation behavior. It provides significant insights into the required forces for generating a fermion-exciton condensate, which will likely be invaluable for realizing such condensations in realistic materials with applications from superconductors to excitonic materials.
\end{abstract}



\maketitle

\section{Introduction}
Model Hamiltonians are theoretical tools that are often useful in simulating the key physics associated with large-scale, highly-correlated systems.  They are capable of modeling an array of quantum phases and many-body phenomena such as phase transitions \cite{Rabe199459,Ostilli2006,Debnath2021,Cai2021,Farias2021}, superconductivity \cite{Richardson_1963,Richardson_1963_2,Richardson_1964,Richardson_1965,BCS1957}, quantum magnetism \cite{Korenblit2012,Sarria2015,Farnell2019180,Sroda2021}, exciton condensation  \cite{Lipkin_model,texpansion,Debergh_2001,David2004,Heiss_2006,Castanos_2006,Co_2018}, lattice-like systems \cite{Erdahl_2000,Sabre_2018},  etc.  Additionally, model Hamiltonians which encompass nontrivial physics are often useful as benchmarks for theoretical tools such as many-body approximations \cite{Richardson_1963,Chu2020,Khamoshi2021,Hu20122157}.

Condensation phenomena---which are inherently highly-correlated---have a long history of being computationally studied through the lens of model Hamiltonians as traditional band theory is inaccurate for such highly-entangled materials \cite{Zheng_2018,Al-Sugheir2016697,Sager_2020,Richardson_1963_2,Richardson_1963,Lipkin_model,BCS1957}.  Specifically, superconductors---materials in which fermion-fermion (Cooper/electron-electron) pairs aggregate into a single quantum state, resulting in the superfluidity of the fermion-fermion pairs---are often explored through use of the Pairing-Force (PF) Hamiltonian \cite{Richardson_1963,Richardson_1963_2,Richardson_1964,Richardson_1965}, which is additionally referred to as the Standard Reduced  Bardeen-Cooper-Schrieffer (BCS) Hamiltonian \cite{BCS1957,Delft_1996,Khamoshi_2020}.  This Hamiltonian is a simple representation of superconductivity as it describes a system with bound Cooper (or Cooper-like particle-particle) pairs interacting in an attractive manner with the holes they leave behind in a Fermi sea with the high-correlation limit of this Hamiltonian resulting in well-known, number-projected BCS wave functions \cite{Richardson_1963_2,Scuseria_BCS}.  Similarly, exciton condensation---in which particle-hole (exciton) pairs condense into a single quantum state resulting in the superfluidity of the composite excitons \cite{keldysh_2017}---can be modeled according to the Lipkin-Meshkov-Glick (LMG) Hamiltonian, which is often simply referred to as the Lipkin model \cite{Lipkin_model,David1998,texpansion,Debergh_2001,David2004,Heiss_2006,Castanos_2006,Co_2018,Sager_2020}.  This Hamiltonian is a highly-degenerate system in which partnered orbitals are inherently particle-hole paired and whose strongly-correlated form results in ground states that demonstrate character of exciton condensation.

Here, we introduce a model Hamiltonian that is capable of capturing fermion-exciton condensation, a new class of highly-correlated condensation phenomena in which both fermion-pair and exciton condensations coexist in a single quantum state.  We demonstrate such coexistent condensate character by calculating the quantum signatures of fermion-pair \cite{Y1962,S1965} and exciton \cite{GR1969,Shiva} condensations (see Sec. \ref{sec:theory} and Appendix \ref{app:signatures}) for systems of even particle numbers ranging from $N=4$ to $N=10$ particles in $r=2N$ orbitals.  These fermion-exciton condensates are shown to be described by wavefunctions which are entanglements of wavefunctions from BCS-like superconductivity and Lipkin-like exciton condensation---consistent with our prior predictions for the large-$N$ thermodynamic limit \cite{Sager_2019} as well as those we observed experimentally on a quantum device \cite{Sager_FEC}.

Our determination of a model Hamiltonian that supports fermion-exciton condensation provides information regarding the nature of the forces necessary to generate such systems---an invaluable first step in the realization of real-world systems that support such dual condensation of excitons and fermion-fermion pairs, which may demonstrate some sort of hybrid of the properties of superconductors and exciton condensates and hence have applications in energy transport and electronics.  The extent of these different phases and the transitions between these phases can also be studied.  Moreover, our Hamiltonian provides an important reference in order to determine whether a given many-body approximation is capable of measuring dual condensate character.

\section{Theory}
\label{sec:theory}
\subsection{Fermion-Pair Condensation}

\begin{figure}[t!]
  \includegraphics[width=8.5cm]{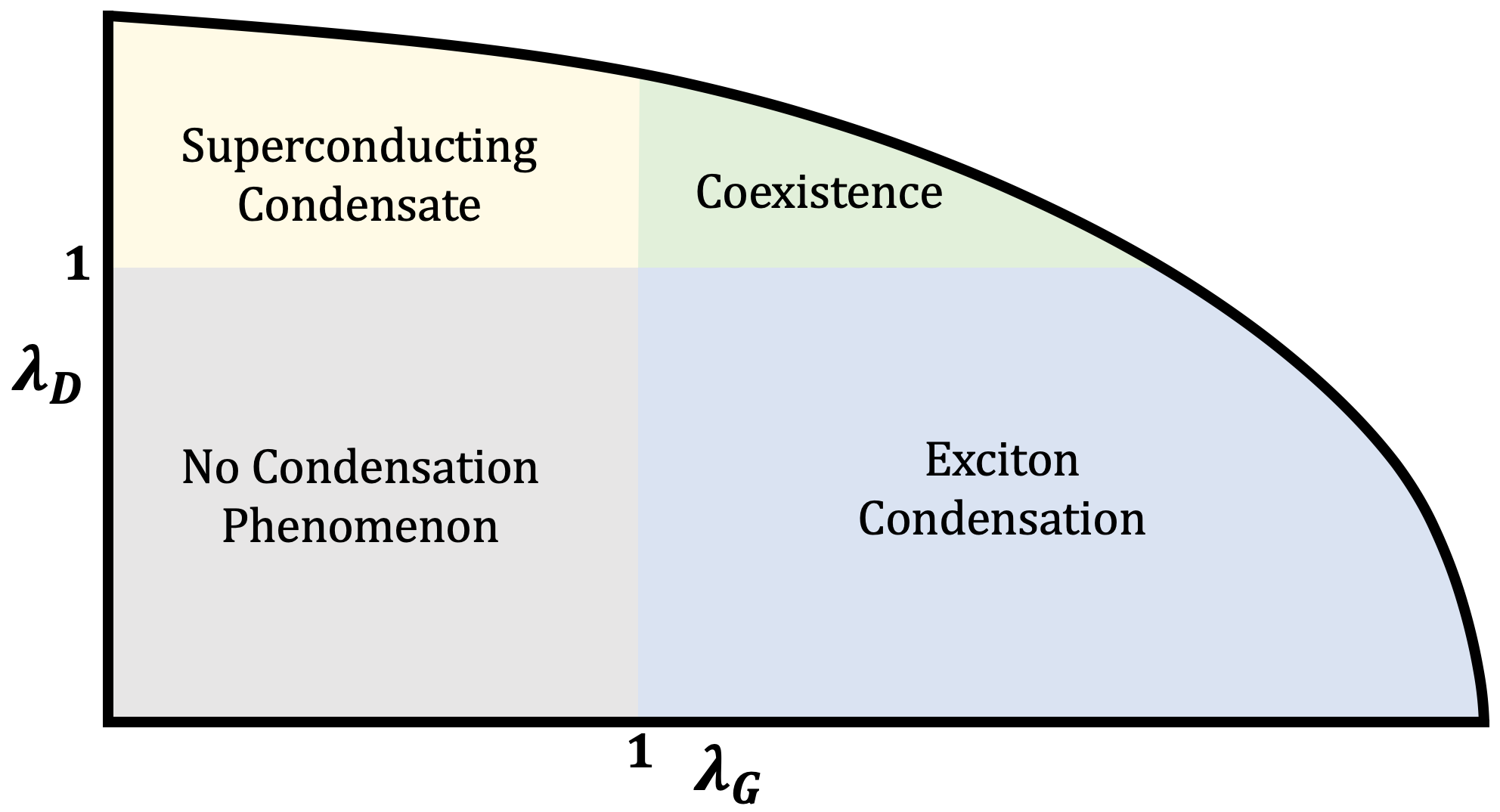}
  \caption{\label{fig:phasediagram}  A figure of the condensate phase diagram in the phase space of the signatures of particle-particle condensation, $\lambda_D$, and exciton condensation, $\lambda_G$, is shown.}
\end{figure}

Superconductivity results from the condensation of bosonic fermion-fermion pairs \cite{bose_einstein_1924,einstein_1924,BCS1957,Anderson_2013}  into a single geminal---a two-fermion function directly analogous to the one-fermion orbital \cite{Y1962,C1963,S1965,RM2015,Srev_1999,shull_1959}---at temperatures below a certain critical temperature.  This condensation of fermion-pairs results in the superfluidity (i.e., frictionless flow) of the constituent particle-particle pairs \cite{london_1938,tisza_1947,BCS1957,Anderson_2013}; if the fermionic pairs are composed of electrons (i.e., Cooper pairs), then these superfluid electron-electron pairs demonstrate superconductivity.

As was first demonstrated by Yang \cite{Y1962} and Sasaki \cite{S1965}, a computational signature of such superconducting states is a large eigenvalue in the particle-particle reduced density matrix (2-RDM), whose elements are given by
\begin{equation}
^{2} D_{k,l}^{i,j} = \langle \Psi | {\hat a}^{\dagger}_i {\hat a}^{\dagger}_j {\hat a}_l {\hat a}_k  | \Psi \rangle
\label{eq:D2}
\end{equation}
where $|\Psi\rangle$ is an $N$-fermion wavefunction and where $\hat{a}_i^\dagger$ and $\hat{a}_i$ are fermionic creation and annihilation operators for orbital $i$, respectively. As eigenvalues of the 2-RDM can be interpreted as the occupations of the two-fermion geminals \cite{Coleman_1965}, when the largest eigenvalue of the 2-RDM---the signature of particle-particle condensation, represented by $\lambda_D$---exceeds the Pauli-like limit of one ($\lambda_D>1$), multiple fermion-fermion pairs occupy a single geminal and hence superconducting character is observed.
This signature is known to directly probe the presence and extent of non-classical (off-diagonal) long-range order \cite{RM2015}.  (See the Appendix for more details on how the signature of superconductivity, $\lambda_D$, was computed.)

The Pairing-Force (PF) model \cite{Richardson_1963,Richardson_1963_2,Richardson_1964,Richardson_1965}---also called the Standard Reduced Bardeen-Cooper-Schrieffer (BCS) model \cite{BCS1957,Delft_1996,Khamoshi_2020}---is known to  exhibit superconducting character in the strong correlation limit and hence achieve a large $\lambda_D$.  The Hamiltonian for the PF model is given in second quantization by
\begin{gather}
\mathcal{H}_{PF}=\frac{1}{2}\sum\limits_{\sigma=\uparrow,\downarrow}\sum\limits_{p=1}^N\epsilon_p\hat{a}^\dagger_{p,\sigma}\hat{a}_{p,\sigma}-G\sum\limits_{p=1}^N\sum\limits_{q=1}^N\hat{a}^\dagger_{p,\uparrow}\hat{a}^\dagger_{p,\downarrow}\hat{a}_{q,\downarrow}\hat{a}_{q,\uparrow}
\label{eq:PFHamiltonian}
\end{gather}
where $p$ is a quantum number that represents a pair of orbitals denoted as $p,\uparrow$ and $p,\downarrow$ with the same energy, where the energies ($\epsilon_p$) are considered to be known, and where the parameter $G$ is a constant that tunes the strength of the pairwise interactions.  Note that in the limit of strong correlation ($G>>\epsilon_p$), maximal superconducting character---$\lambda_{D} = \frac{N}{2} \left ( 1-\frac{N-2}{r} \right )$ \cite{C1963,Coleman_1965}---is observed.

\subsection{Exciton Condensation}
Directly analogous to superconductivity resulting from bosonic particle-particle pairs condensing into a single particle-particle function, exciton condensation results from the condensation of particle-hole pairs (i.e., excitons) into a single particle-hole function below a certain critical temperature, which results in the superfluidity of the excitons \cite{Fil_Shevchenko_Rev,keldysh_2017}. Exciton condensates, while difficult to realize experimentally, have been observed in systems composed of polaritons (excitons coupled to photons) \cite{KRK2006,fuhrer_hamilton_2016,YZ_2021} and in two-dimensional structures such as semiconductors \cite{Butov1994}, graphene bilayers \cite{Liu2017, Min2008,Cao_2018}, and van der Waals heterostructures \cite{Sigl2020,Wang2019,Fogler2014, Kogar2017}.

The signature of exciton condensation---denoted as $\lambda_G$---is similarly analogous to that for fermion-pair condensation; the presence and extent of exciton condensate character can be measured from the largest eigenvalue of a modified particle-hole reduced density matrix given by \cite{Shiva, GR1969, Kohn1970}
\begin{multline}
{}^{2}\Tilde{G}^{i,j}_{k,l}={}^{2}G^{i,j}_{k,l}-{}^{1}D^i_k{}^{1}D^j_l \\=\langle \Psi | {\hat a}^{\dagger}_i {\hat a}_j {\hat a}^{\dagger}_l{\hat a}_k  | \Psi \rangle-\langle\Psi|\hat{a}^\dagger_i\hat{a}_k|\Psi\rangle\langle\Psi|\hat{a}^\dagger_j\hat{a}_l|\Psi\rangle
\label{eq:modG2}
\end{multline}
where ${}^{1}D$ is the one-fermion reduced density matrix (1-RDM).  Note that this modification removes the extraneous large eigenvalue from a ground-state-to-ground-state transition such that a signature above one ($\lambda_G>1$) is indicative of exciton condensation. (See the Appendix for more details on how the signature of exciton condensation, $\lambda_G$ was computed.) This computational signature has been utilized to study exciton condensation is possible in quantum and molecular systems \cite{Shiva,Sager_2019,Sager_2020,Sager_FEC,Anna_2021}.

One model known to achieve a large $\lambda_G$ value and hence exhibit exciton condensate character in the limit of a large correlation is the Lipkin quasispin model \cite{Lipkin_model,texpansion,Debergh_2001,David2004,Heiss_2006,Castanos_2006,Co_2018}.
The $N$-fermion Lipkin quasispin model consists of two energy levels $\left\{-\frac{\epsilon}{2},\frac{\epsilon}{2}\right\}$, each containing $N$ energetically-degenerate states.  The second-quantized Hamiltonian can be expressed as   \cite{David2004}
\begin{gather}
\mathcal{H}_{L}=\frac{\epsilon}{2}\sum\limits_{\sigma=\pm1}\sigma\sum\limits_{p=1}^N\hat{a}^\dagger_{\sigma,p}\hat{a}_{\sigma,p}\nonumber\\+\frac{\gamma}{2}\sum\limits_{\sigma=\pm 1}\sum\limits_{p,q=1}^N\hat{a}^\dagger_{+\sigma,p}\hat{a}_{-\sigma,p}\hat{a}^\dagger_{-\sigma,q}\hat{a}_{+\sigma,q}\nonumber\\+\frac{\lambda}{2}\sum\limits_{\sigma=\pm1}\sum\limits_{p,q=1}^N \hat{a}^\dagger_{+\sigma,p}\hat{a}^\dagger_{+\sigma,q}\hat{a}_{-\sigma,q}\hat{a}_{-\sigma,p}
\label{eq:LipkinHam}
\end{gather}
where $\sigma=\pm 1$ and $p=1,2,\dots,N$ are quantum numbers that completely characterize the system in which $p$ describes the site number labelling the $N$ states in a given level and $\sigma$ represents the upper ($+1$) or lower ($-1$) energy levels, respectively.  Note that in this model, the $\lambda$ term allows for double excitations and de-excitations, and the $\gamma$ term allows for a single particle to be scattered up while another is simultaneously scattered down; as a result, in the Lipkin model, only even excitations are allowed, and only one particle may occupy a given site (i.e., have a specific quantum number $p$) such that each site in the lower level is particle-hole paired with the corresponding site in the upper level.  By having the terms correlating orbitals in the Hamiltonian ($\lambda,\gamma$) be sufficiently larger than the energy term (i.e., in the limit of high correlation), maximal exciton condensation---$\lambda_G=\frac{N}{2}$ \cite{GR1969}---can be obtained for $\lambda=\gamma$.

\subsection{Fermion-Exciton Condensation}
A fermion-exciton condensate is a single quantum state that simultaneously demonstrates character of superconductivity and exciton condensation, i.e., both signatures of condensation---the largest eigenvalue of the particle-particle RDM (Eq. (\ref{eq:D2})) and the largest eigenvalue of the modified particle-hole RDM (Eq. (\ref{eq:modG2}))---are simultaneously large ($\lambda_D,\lambda_G>1$). \cite{Sager_2019}.

\begin{figure}[tbh!]
\centering
\includegraphics[width=6cm]{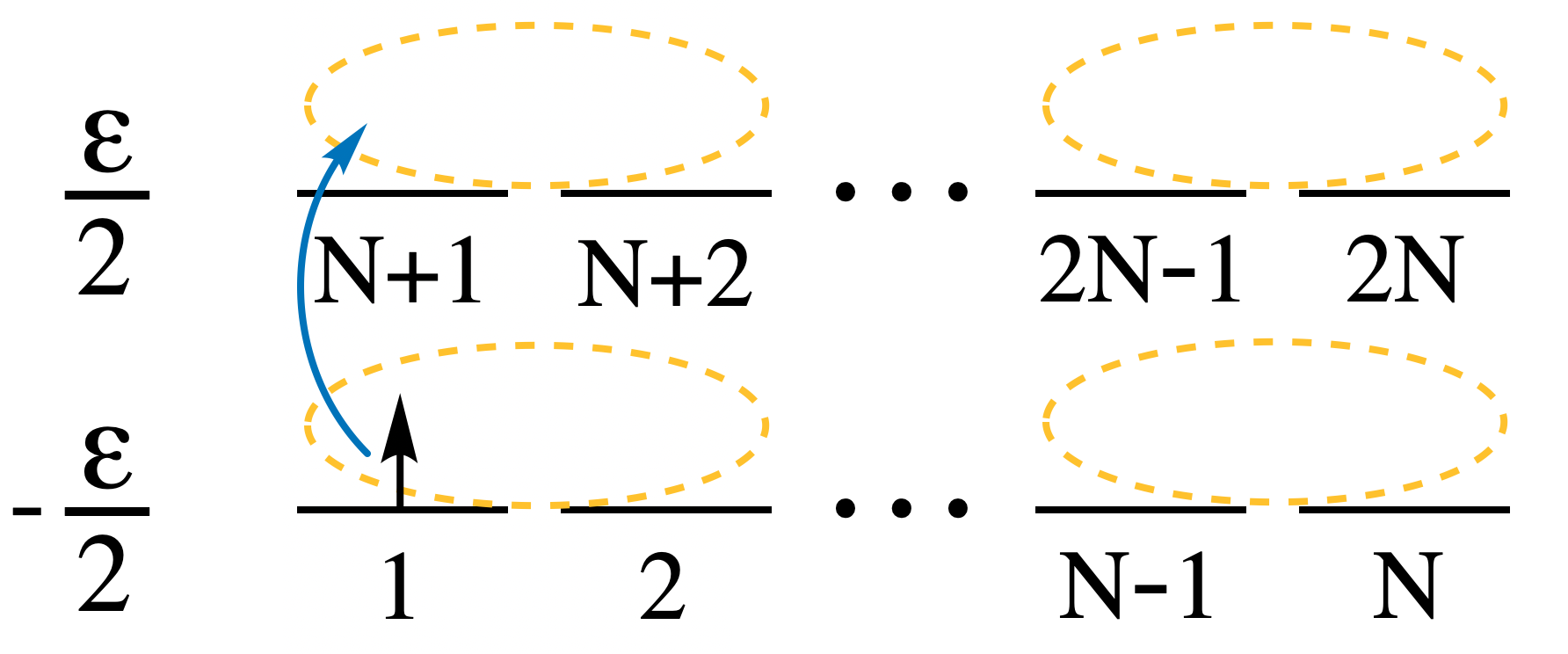}
\caption{\label{fig:model} A pictorial representation of the model Hamiltonian we introduce in which there are two $N$-degenerate energy levels---with energies $-\frac{\epsilon}{2}$ and $\frac{\epsilon}{2}$---with double excitations and de-excitations, scattering in which one particle is de-excited while another is simultaneously excited, and a pair-wise interaction term between sites $2j-1$ and $2j$ for $j\in \{1,2,\dots,N\}$ (yellow circles) is shown.  Note that the Lipkin-like excitations must occur within a site ($p\leftrightarrow p+N$, blue arrow).}
\end{figure}

To gain insight into such fermion-exciton condensates, here we propose a model system that is capable of demonstrating simultaneous fermion-pair and exciton condensate character.  In this model, we introduce the pairwise interaction from the Pairing-Force model into the scaffolding of the Lipkin model; thus, the model keeps the structure of the Lipkin model in which $N$ particles occupy two $N$-degenerate energy levels ($-\frac{\epsilon}{2}$ and $\frac{\epsilon}{2}$) with allowed double excitations on two sites ($\lambda$) and simultaneous scattering of a particle up on one site and down on another ($\gamma$)---where Lipkin-like sites are now given as orbitals $p$ and $p+N$; however, we additionally pair adjacent orbitals---orbitals $2j-1$ and $2j$ for $j\in\{1,2,\dots,N\}$---via the PF parameter, $G$. (See Fig. \ref{fig:model}.)  The Hamiltonian for this model is thus given by
\begin{gather}
\mathcal{H}=-\frac{\epsilon}{2}\sum\limits_{i=1}^N\hat{a}^\dagger_i\hat{a}_i+\frac{\epsilon}{2}\sum\limits_{i=1}^N\hat{a}^\dagger_{i+N}\hat{a}_{i+N}\nonumber\\+\frac{\lambda}{2}\sum\limits_{p=1}^N\sum\limits_{q=1}^N\hat{a}^\dagger_p\hat{a}^\dagger_{q}\hat{a}_{q+N}\hat{a}_{p+N}+\frac{\lambda}{2}\sum\limits_{p=1}^N\sum\limits_{q=1}^N\hat{a}^\dagger_{p+N}\hat{a}^\dagger_{q+N}\hat{a}_{q}\hat{a}_{p}\nonumber\\+\frac{\gamma}{2}\sum\limits_{p=1}^N\sum\limits_{q=1}^N\hat{a}^\dagger_{p+N}\hat{a}^\dagger_{q}\hat{a}_{q+N}\hat{a}_{p}+\frac{\gamma}{2}\sum\limits_{p=1}^N\sum\limits_{q=1}^N\hat{a}^\dagger_{p}\hat{a}^\dagger_{q+N}\hat{a}_{q}\hat{a}_{p+N}\nonumber\\-G\sum\limits_{j=1}^N\sum\limits_{k=1}^N\hat{a}^\dagger_{2j-1}\hat{a}^\dagger_{2j}\hat{a}_{2k}\hat{a}_{2k-1}
\label{eq:Model}
\end{gather}
in second quantization, with a given set of parameters ($\epsilon,\lambda,\gamma,G$) directly determining the extent of fermion-pair and exciton condensation ($\lambda_D$ and $\lambda_G$, respectively) of the ground state corresponding to this model Hamiltonian.

While this model Hamiltonian is not the first to combine the pairwise interaction from the Pairing-Force model with the Lipkin model, the model Hamiltonian introduced by Plastino and coworkers causes direct competition between particle-particle and particle-hole correlations and hence proves incapable of demonstrating a fermion-exciton condensate phase (see Appendix \ref{app:plastino}) \cite{Plastino_1978,Plastino_2018,Plastino_2021}.  Conversely, due to our introduction of the Pairing-Force interactions between adjacent orbitals instead of orbitals in the same Lipkin-like site, particle-particle and particle hole pairing can coexist and hence fermion-pair-exciton (FEC) states can be achieved as is shown in the results that follow.

\section{Results}
\subsection{$N=4$, The Minimal FEC}
\begin{figure}[tbh!]
\centering
\includegraphics[width = 6cm]{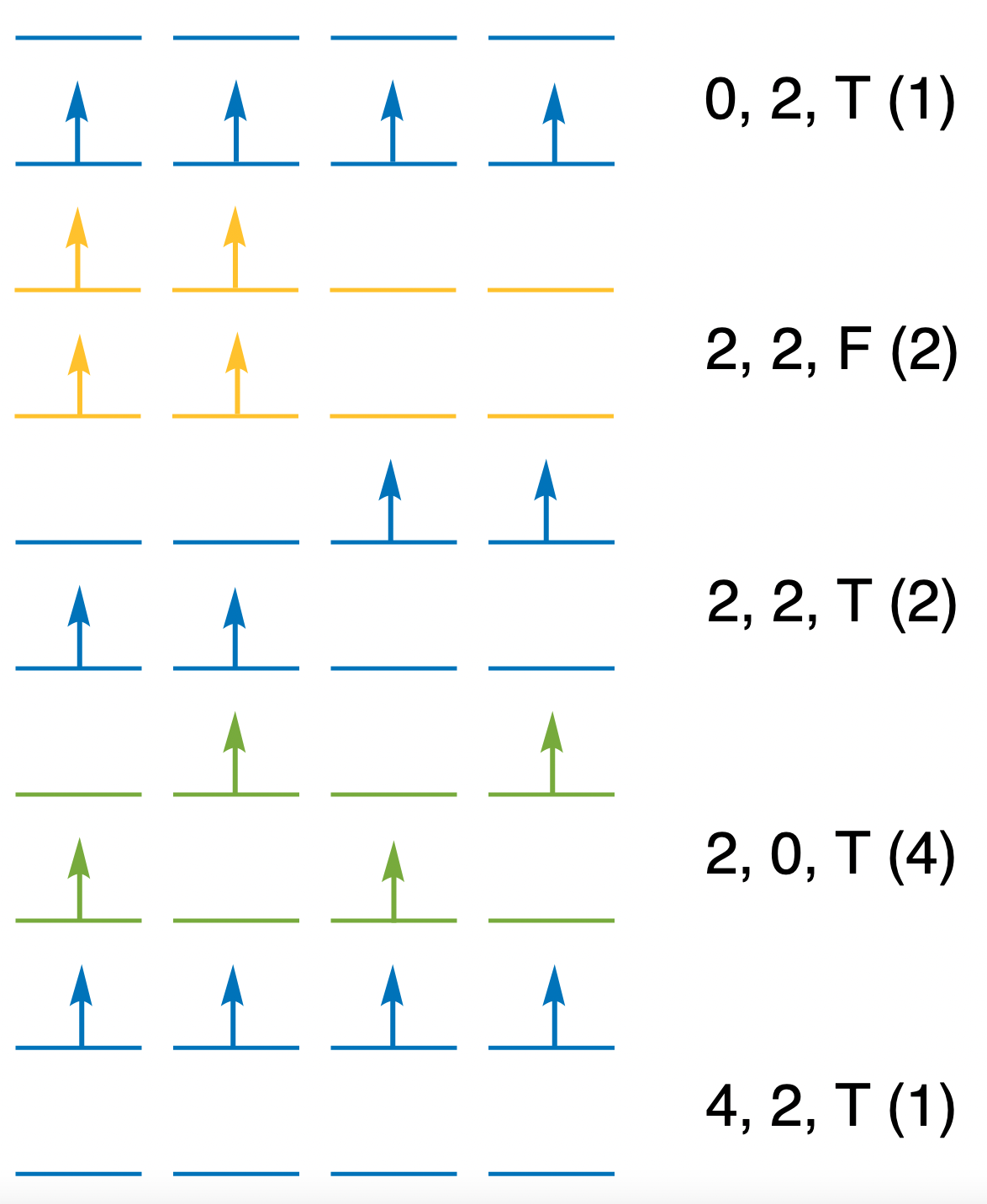}
\caption{\label{fig:N4arrows} Configurations representing each of the five classes of non-zero basis states for the FEC Hamiltonian for $N,r=4,8$ are shown where each label $x,y,bool$ represents the number of particles excited to the upper $N$-degenerate energy level ($x$), the number of BCS-like pairs ($y$), and whether the configuration is consistent with the Lipkin model ($bool$), where the degeneracy of each class of states is given in parenthesis, and where green, yellow, and blue configurations represent that the corresponding states are consistent with only the Lipkin Hamiltonian, only the Pairing-Force Hamiltonian, or both Lipkin and PF Hamiltonians, respectively.}
\end{figure}

As the authors have previously demonstrated \cite{Sager_2019}, a system with as few as $N=4$ particles in $r=8$ orbitals can support formation of a fermion-exciton condensate.  As such, we first fully explore such a minimalistic FEC system. The ground state of the FEC Hamiltonian that we have introduced---Equation (\ref{eq:Model})---for four particles has contributions from only ten of the seventy ($rCN$) possible configurations.  Of these ten basis states, there are only five distinct classes composed of degenerate orientations---see Fig. \ref{fig:N4arrows}---that allow for the direct computation of a matrix-form of the Hamiltonian in a minimal basis state.  The five basis states are defined by three quantum numbers, $x,y,bool$, where the first indicates the number of particles excited to the upper energy level ($x$), the second indicates the number of BCS-like pairs (number of times both $2j-1$ and $2j$ are occupied, $y$), and the third is a boolean that indicates whether the configuration is ``Lipkin''-like in the regard that no two orbitals representing a ``Lipkin'' site (denoted as $p$ and $p+N$, see the blue arrow in Fig. \ref{fig:N4arrows}) are dually occupied or dually unoccupied.

Utilizing the basis shown in Fig. \ref{fig:N4arrows}---$|0,2,T\rangle, \ |2,2,F\rangle, \ |2,2,T\rangle, \ |2,0,T\rangle,$ and $|4,2,T\rangle$---the Hamiltonian from Eq. (\ref{eq:Model}) can be represented by \begin{equation}
\mathcal{H}_4=\left(\begin{array}{ccccc}
  -2\epsilon-2G & -G\sqrt{2} & \frac{2\lambda-2G}{\sqrt{2}} &  2\lambda & 0 \\
  -G\sqrt{2} & -2G+2\gamma & -2G & 0 & -G\sqrt{2} \\
   \frac{2\lambda-2G}{\sqrt{2}} & -2G & -2G & 2\gamma\sqrt{2} &  \frac{2\lambda-2G}{\sqrt{2}} \\
  2\lambda & 0 & 2\gamma\sqrt{2} & 2\gamma & 2\lambda \\
  0 & -G\sqrt{2} &  \frac{2\lambda-2G}{\sqrt{2}} & 2\lambda & 2\epsilon-2G
\end{array}\right)
\label{eq:Ham4}
\end{equation}
where each term---corresponding to the interaction between two classes of basis states, $|i\rangle$ and $|j\rangle$---is obtained from programmatically generating all sets of second-quantization creation and annihilation operators in Eq. (\ref{eq:Model}), taking the expectation value for each combination of pairs of configurations in classes $|i\rangle$ and $|j\rangle$, summing the results, and normalizing by dividing by the square root of the number of configurations for both $|i\rangle$ and $|j\rangle$.  For example, if $|i\rangle=|2,2,F\rangle=\left(|1,2,5,6\rangle+|3,4,7,8\rangle\right)/\sqrt{2}$ and $|j\rangle=|2,2,T\rangle=\left(|1,2,7,8\rangle+|3,4,5,6\rangle\right)/\sqrt{2}$, the Hamiltonian term would be given by
\begin{gather}
\frac{\left(\langle1,2,5,6|+\langle3,4,7,8|\right)}{\sqrt{2}}\mathcal{H}_4\frac{\left(|1,2,7,8\rangle+|3,4,5,6\rangle\right)}{\sqrt{2}} \nonumber\\
=\frac{1}{2}[\langle 1,2,5,6|\mathcal{H}_4|1,2,7,8\rangle+\langle 1,2,5,6|\mathcal{H}_4|3,4,5,6\rangle\nonumber\\+\langle 3,4,7,8|\mathcal{H}_4|1,2,7,8\rangle+\langle 3,4,7,8|\mathcal{H}_4|3,4,5,6\rangle]
\label{eq:Ham4Ex}
\end{gather}

\begin{figure*}[tb!]
\subfloat[$N,r=4,8$]{\label{fig:N4plot}\includegraphics[width = 8cm]{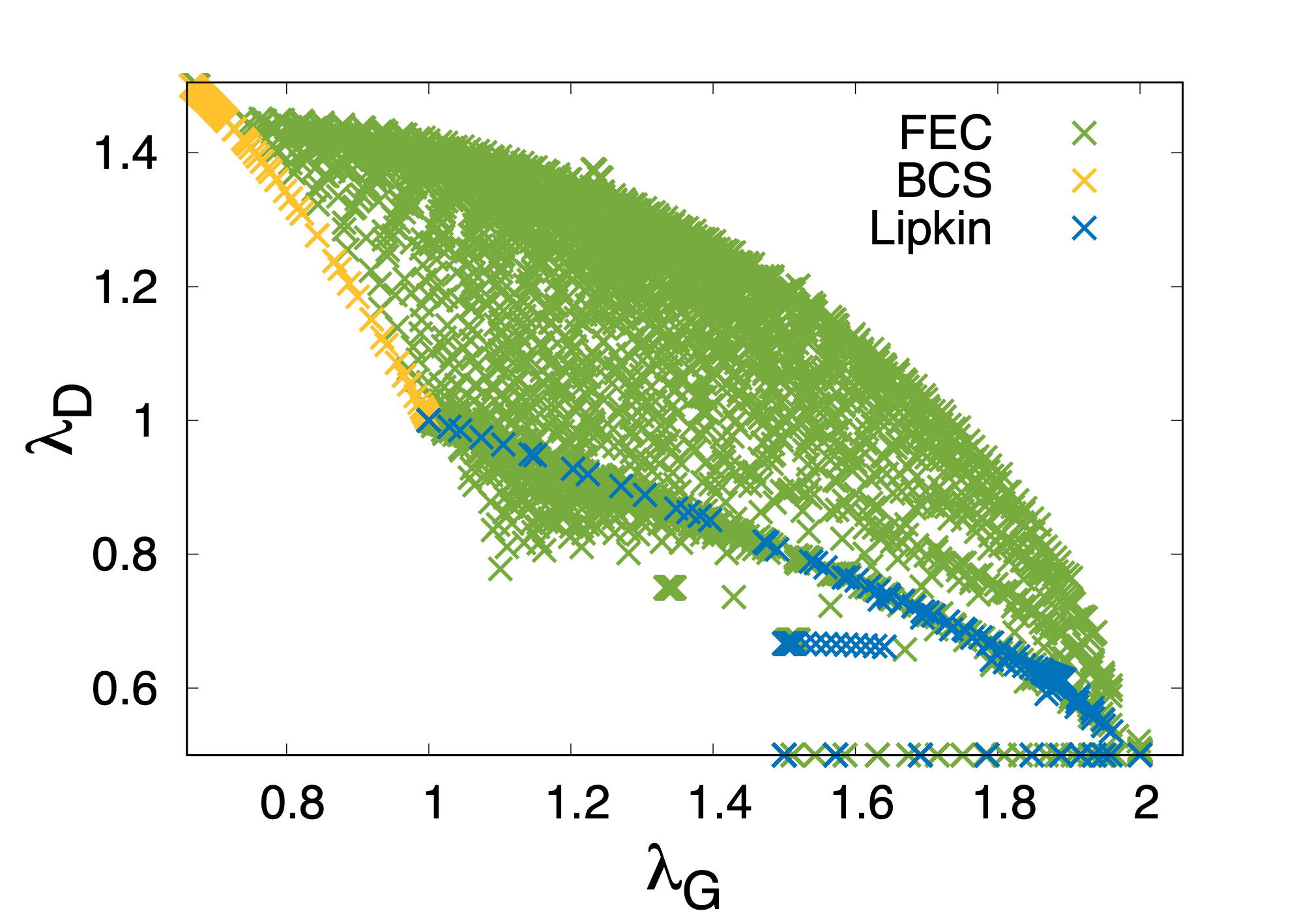}}
\subfloat[$N,r=6,12$]{\label{fig:N6plot}\includegraphics[width = 8cm]{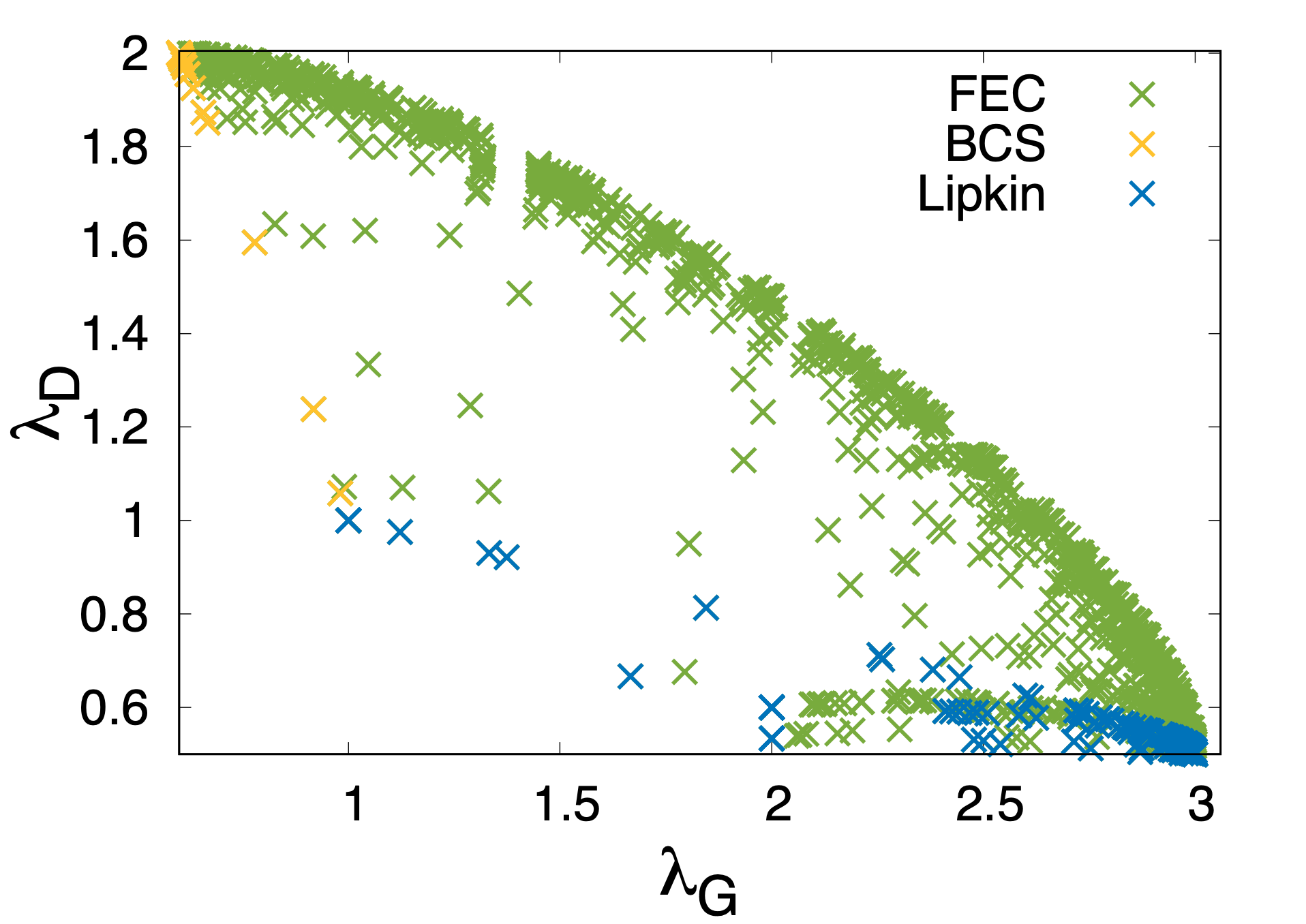}}\\ \subfloat[$N,r=8,16$]{\label{fig:N8plot}\includegraphics[width = 8cm]{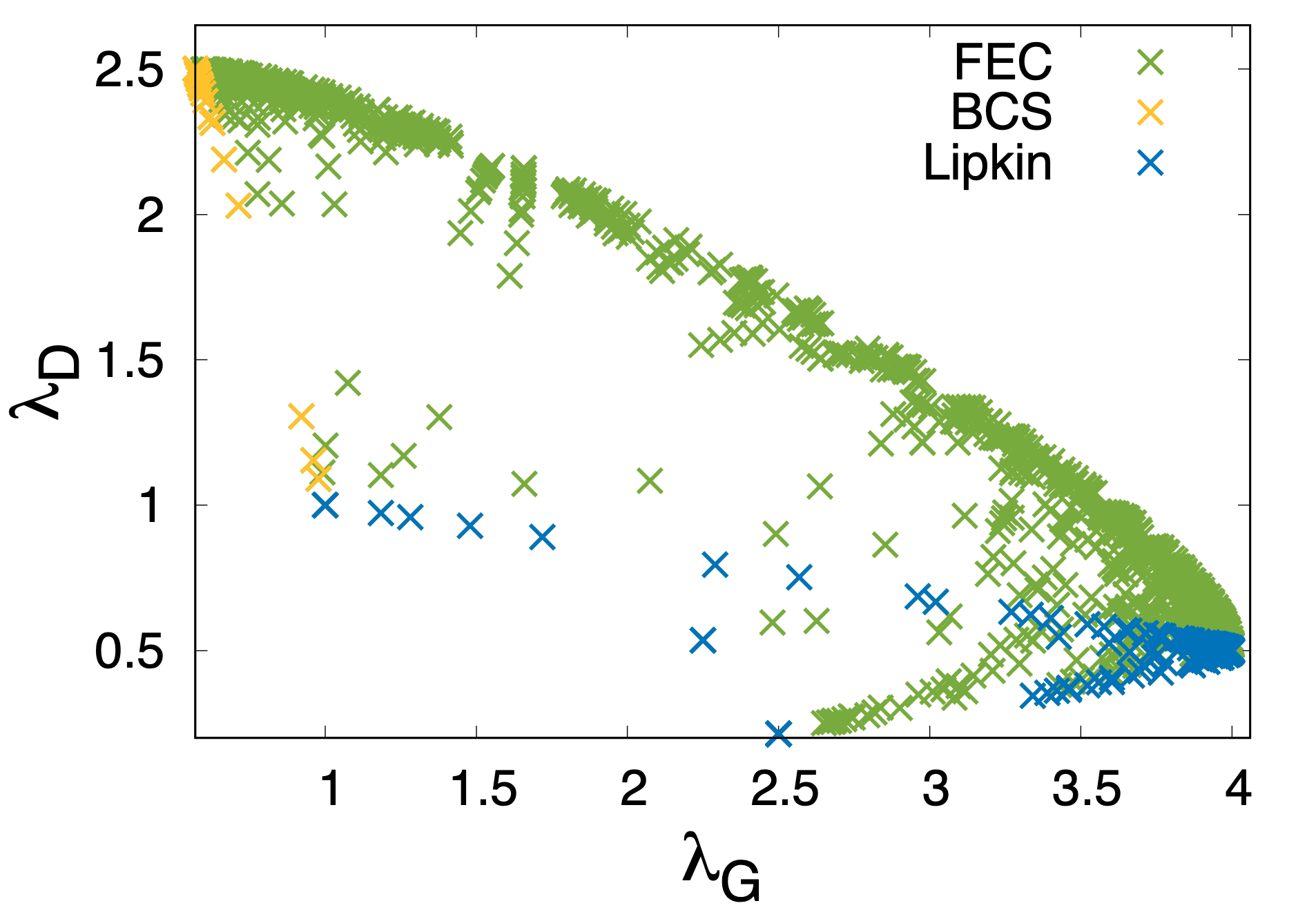}}
\subfloat[$N,r=10,20$]{\label{fig:N10plot}\includegraphics[width = 8cm]{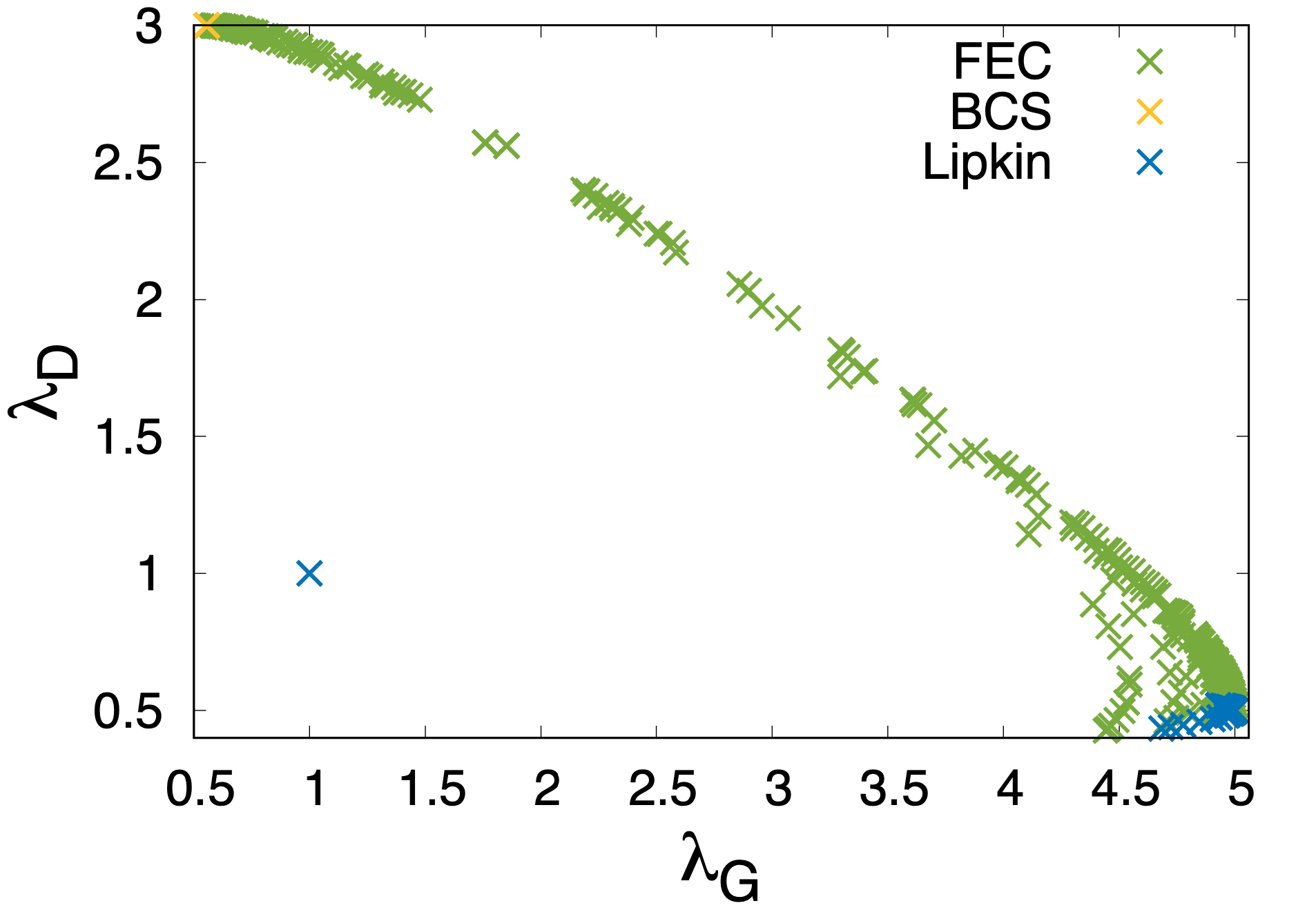}}
\caption{\label{fig:Nplots} Plots of $\lambda_G$ versus $\lambda_D$ where parameters in the FEC Hamiltonian are systematically varied are shown for systems involving (a) $N=4$, (b) $N=6$, (c) $N=8$, and (d) $N=10$ particles in $r=2N$ orbitals.}
\end{figure*}

Fig. \ref{fig:N4plot} scans over the signatures of condensation---$\lambda_D$ and $\lambda_G$---for the ground state of the Hamiltonian in Eq. (\ref{eq:Ham4}) by systematically varying the parameters $\epsilon$, $\lambda$, $\gamma$, and $G$ where the yellow BCS x's represent ground states in which the PF Hamiltonian is implemented (i.e., $\lambda=\gamma=0$), the blue Lipkin x's represent states in which the Lipkin Hamiltonian is implemented (i.e., $G=0$), and where the green FEC x's represent states with character of both PF and Lipkin Hamiltonians.  As this figure demonstrates, the largest degree of superconducting character (the largest $\lambda_D$) is indeed observed in the BCS limit of the FEC Hamiltonian (when $G>>\epsilon, \ \lambda=\gamma\approx0$), and the largest degree of exciton condensate character (the largest $\lambda_G$) is observed in the Lipkin limit of the FEC Hamiltonian ($\lambda\approx\gamma>>\epsilon, \ G\approx0$).  However, neither the BCS nor Lipkin limits of the Hamiltonian is capable of demonstrating a dual fermion-exciton condensate as $\lambda_D$ and $\lambda_G$ only simultaneously exceed the Pauli-like limit of one when the full FEC Hamiltonian from Eq. (\ref{eq:Model}) is implemented including both BCS-like ($G$) and Lipkin-like ($\lambda,\gamma$) terms.

Our model FEC Hamiltonian, however, is capable of demonstrating a wide variety of dual condensate character as a variety of input parameters lead to ground state configurations in which both $\lambda_G$ and $\lambda_D$ simultaneously exceed one.
Additionally, the $\lambda_D$ and $\lambda_G$ values obtained by scanning over the Hamiltonian parameters (in Fig. \ref{fig:N4plot}) demonstrate an elliptic nature consistent with the convex nature of 2-RDMs projected onto a two-dimensional space \cite{schwerdtfeger_mazziotti_2009,gidofalvi_mazziotti_2006,zauner_2016} that matches predictions for a FEC that these authors first presented in Ref. \onlinecite{Sager_2019}.  This elliptic boundary as well as the density of points in the zone corresponding to fermion-exciton condensate character indicate that the FEC model Hamiltonian introduced here is capable of spanning the entirety of the FEC region of $\lambda_D$ versus $\lambda_G$ space (i.e., $\lambda_D,\lambda_G>1$).

\begin{figure}[tbh!]
\centering
\includegraphics[width = 8cm]{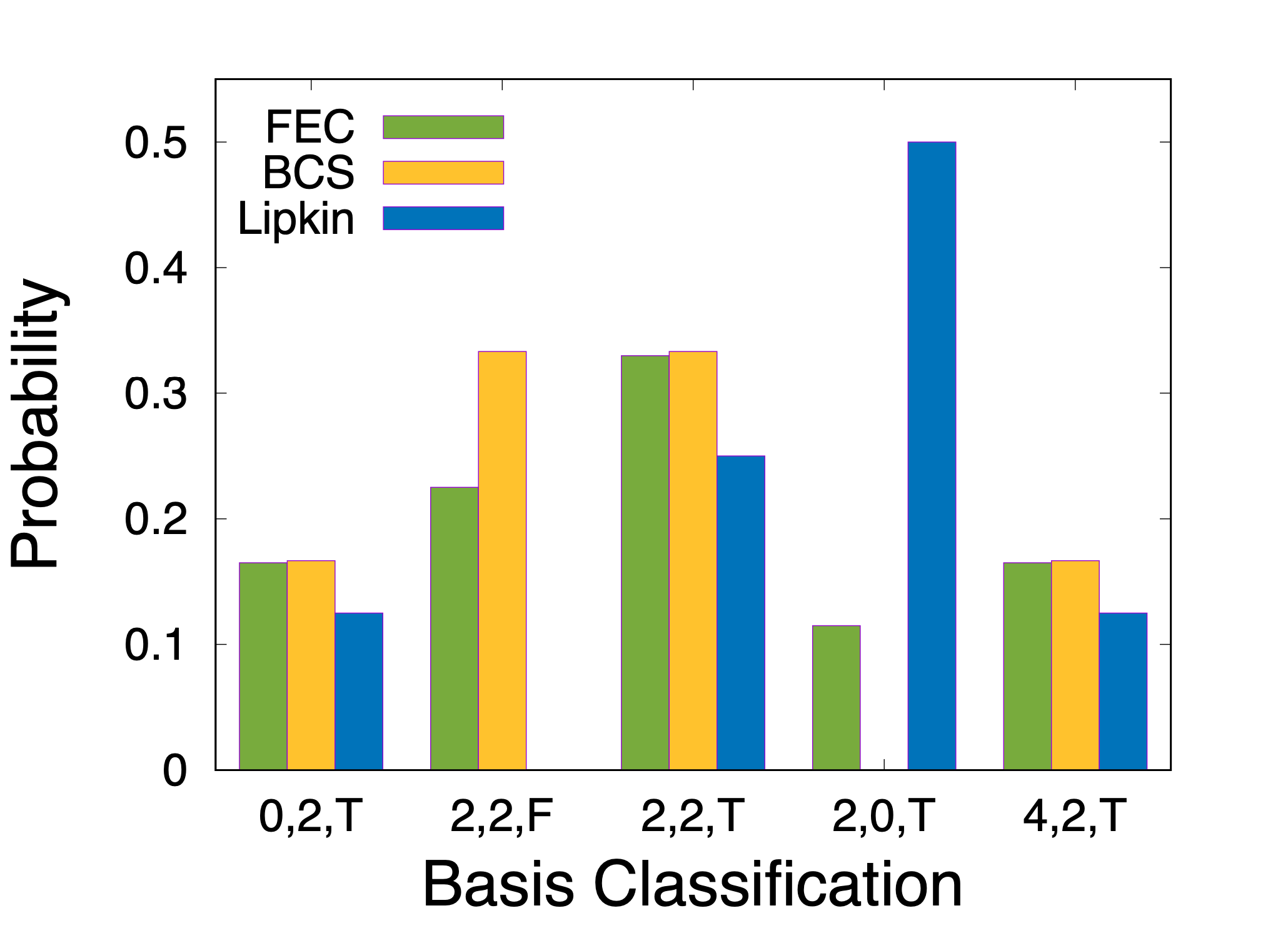}
\caption{\label{fig:N4hist} The probabilities corresponding to each of the five classes of basis states (see Fig. \ref{fig:N4arrows}) consistent with the FEC Hamiltonian for $N,r=4,8$ are shown where green, yellow, and blue bars correspond to the lowest eigenstate of the Lipkin Hamiltonian, the Pairing-Force Hamiltonian, and FEC Hamiltonian, respectively.}
\end{figure}

In Ref. \onlinecite{Sager_2019}, these authors theoretically establish that in the thermodynamic limit, a possible wavefunction demonstrating fermion-exciton condensation can be obtained by entangling wavefunctions that separately demonstrate superconducting character ($|\Psi_D\rangle$ with large $\lambda_D$) and exciton condensate character ($|\Psi_G\rangle$ with large $\lambda_G$) according to
\begin{equation}
    |\Psi_{FEC}\rangle=\frac{1}{\sqrt{2 - |\Delta|}}\left(|\Psi_D\rangle - {\rm sgn}(\Delta) |\Psi_G\rangle \right),
\label{eq:FEC}
\end{equation}
where $\Delta = 2\langle \Psi_D | \Psi_G \rangle$.
In Fig. \ref{fig:N4hist} occupation probabilities for each of the five classes of basis states consistent with the $N,r=4,8$ FEC Hamiltonian that contribute to a BCS wavefunction (yellow, $\epsilon,\lambda,\gamma,G=0,0,0,0.7$, $\lambda_D=1.50$, $\lambda_G=0.67$), a Lipkin wavefunction  (blue, $\epsilon,\lambda,\gamma,G=0,-0.5,-0.5,0$, $\lambda_D=0.50$, $\lambda_G=2.00$), and a FEC wavefunction  (green, $\epsilon,\lambda,\gamma,G=0,-0.5,-0.5,0.7$, $\lambda_D=1.31$, $\lambda_G=1.32$) are given.  From this data, it can be observed that the FEC wavefunction does indeed appear to be an entanglement of the individual BCS and Lipkin wavefunctions for the case of $N=4$; this is consistent with the theoretical result in the thermodynamic limit.

\subsection{Higher-Particle FECs}
In order to observe trends related to system size, we employ the methodologies used to explore the $N,r=4,8$ model system and extrapolate to systems composed of $N=6,8,10$ particles in $r=12,16,20$ orbitals.  Figures summarizing the signatures of superconducting character ($\lambda_D$) and exciton condensate character ($\lambda_G$) obtained for the ground state wavefunctions of these larger model Hamiltonians can be seen in Figs. \ref{fig:N6plot}-\ref{fig:N10plot}.  Similar to the results from the $N=4$ data, elliptic fits spanning the maximal signature of superconducting character observed for the BCS wavefunction to the maximal signature of exciton condensate character for the Lipkin wavefunction with a large variety of parameters supporting dual fermion-exciton condensation.  Note that as the size of the system increases from $N=6$ to $N=8$ to $N=10$, the number of classes of degenerate, non-zero basis states as well as the number of basis states composing each class increase from $8$ classes with a total of $44$ non-zero basis states to $14$ classes with a total of $230$ non-zero basis states to $20$ classes with a total of $1212$ non-zero basis states.  As such, the relative sparsity of the computations in $\lambda_D$ versus $\lambda_G$ as system size is increased is due to fewer computations being run with larger increments between each of the parameters as they are varied.

\begin{figure}[tbh!]
\centering
\includegraphics[width = 8cm]{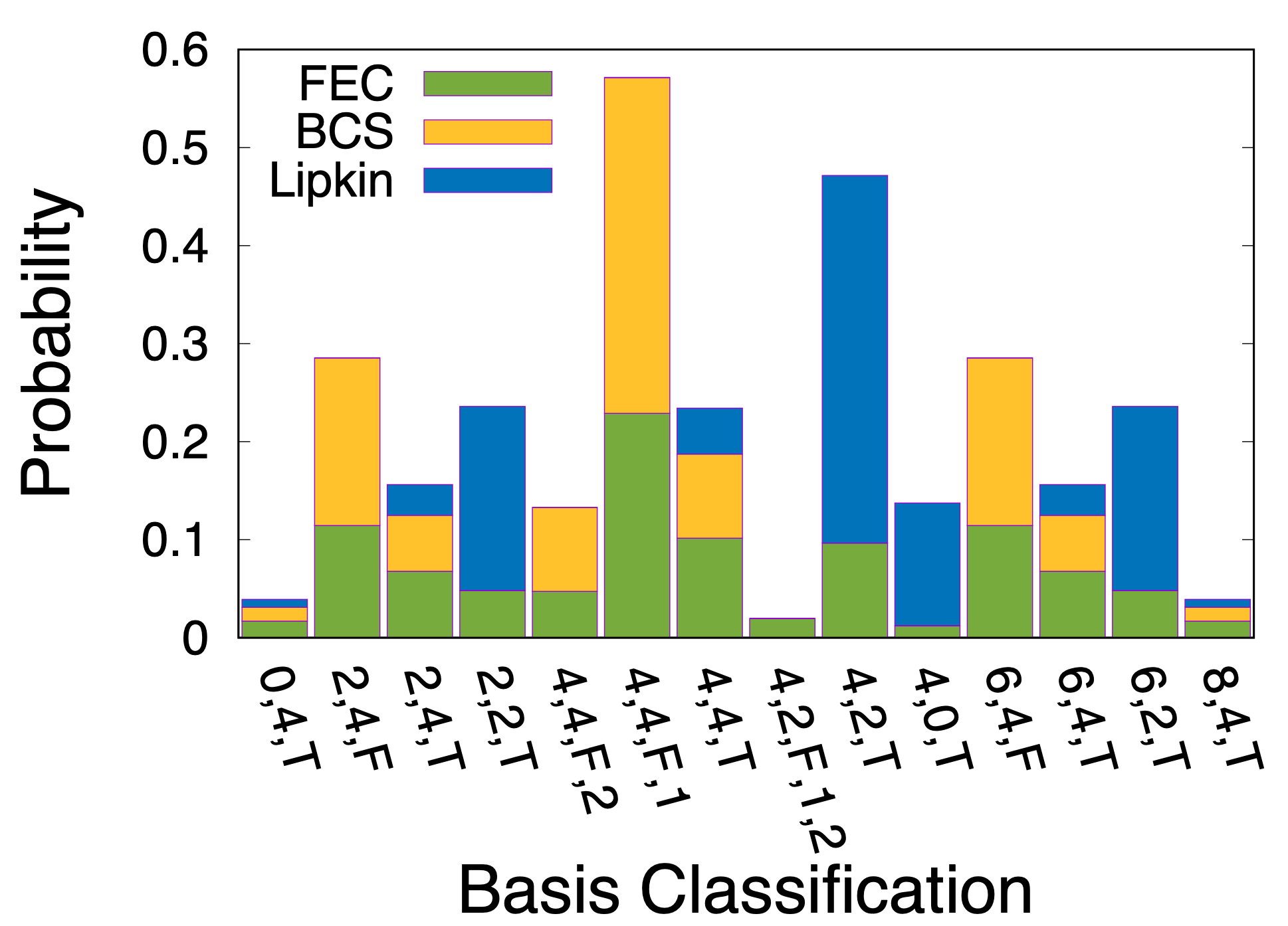}
\caption{\label{fig:N8hist} The probabilities corresponding to each of the fourteen classes of basis states consistent with the FEC Hamiltonian for $N,r=8,16$ are shown where green, yellow, and blue bars correspond to the lowest eigenstate of the Lipkin Hamiltonian, the Pairing-Force Hamiltonian, and FEC Hamiltonian, respectively. Each label $x,y,bool,\zeta,\tau$ represents the number of particles excited to the upper $N$-degenerate energy level ($x$), the number of BCS-like pairs ($y$), whether the configuration is consistent with the Lipkin model ($bool$), the number of times BCS-like pairs are ``stacked'' into the same site ($\zeta$), and the number of times a diagonal configuration occur in which either $2j-1$/$2j+N$ or $2j-1+N$/$2j$ are simultaneously occupied where $2j-1$ and $2j$ are adjacent, paired orbitals ($\tau$).  These values act as quantum numbers that define the degenerate classes of non-zero basis functions composing the ground state to the FEC Hamiltonian.}
\end{figure}

\begin{figure}[tbh!]
\centering
\includegraphics[width = 8cm]{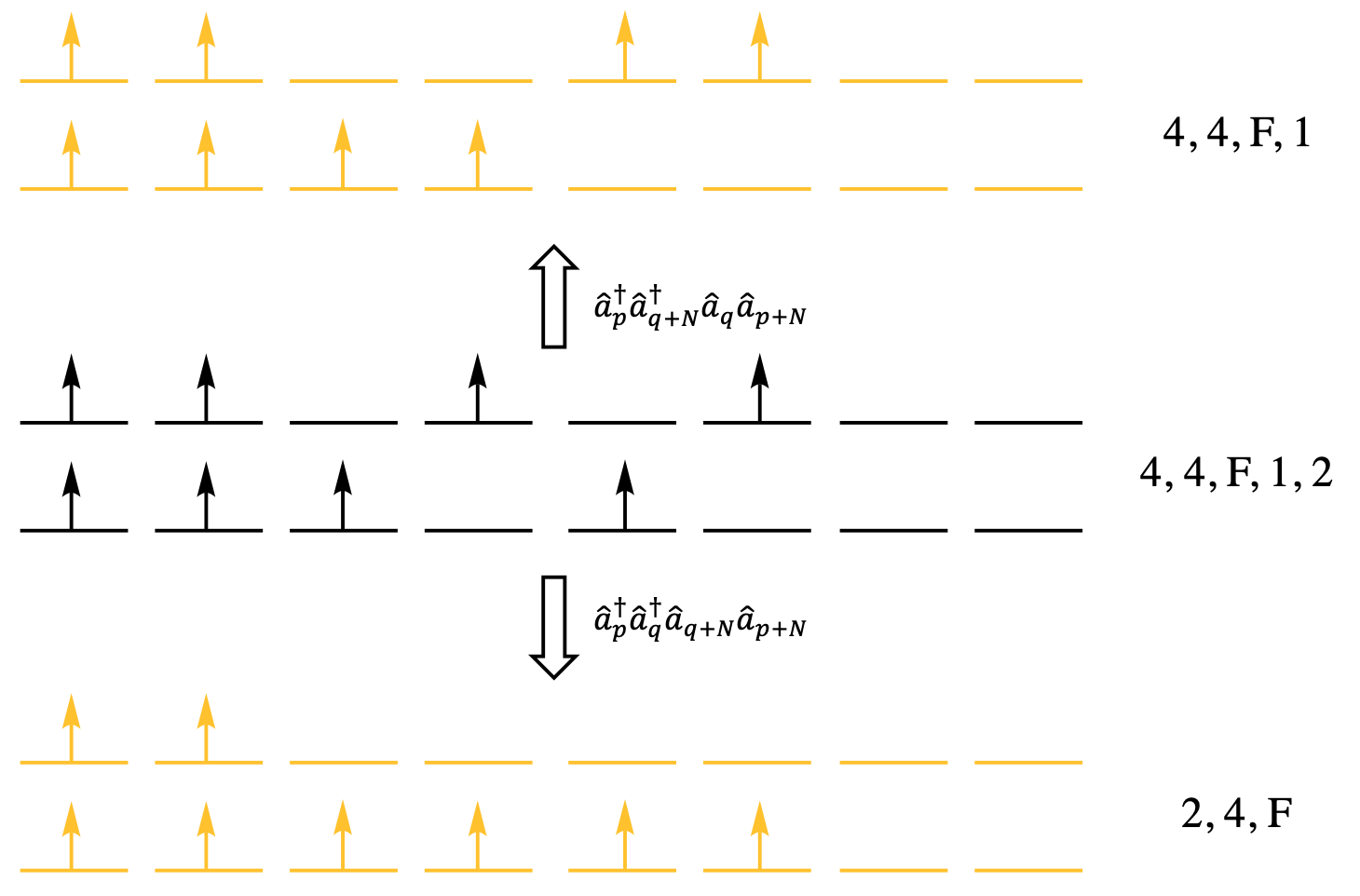}
\caption{\label{fig:N8arrows} Configurations representing how the Lipkin-like double excitation term ($\lambda$) and scattering term ($\gamma$) in the FEC Hamiltonian relate the $|4,4,F,1,2\rangle$ basis state for $N,r=8,16$ to BCS-like basis states.}
\end{figure}

To demonstrate how the classes of non-zero basis states vary as system size is increased, Fig. \ref{fig:N8hist}---which shows the occupation probabilities for each of the fourteen classes of basis states consistent with the $N,r=8,16$ FEC Hamiltonian that contribute to a BCS wavefunction (yellow, $\epsilon,\lambda,\gamma,G=0,0,0,0.9$, $\lambda_D=2.50$, $\lambda_G=0.57$), a Lipkin wavefunction  (blue, $\epsilon,\lambda,\gamma,G=0,-0.5,-0.5,0$, $\lambda_D=0.50$, $\lambda_G=4.00$), and a FEC wavefunction  (green, $\epsilon,\lambda,\gamma,G=0,-0.5,-0.5,0.9$, $\lambda_D=2.06$, $\lambda_G=1.87$)---is included.  Note that due to an increase in the possible complexity, two more quantum numbers are added to describe a few of the classes of basis states; specifically, $\zeta$ and $\tau$ are added to $x,y,bool,\zeta,\tau$ where $\zeta$ corresponds to the number of times BCS-like pairs are ``stacked'' into the same site such that orbitals $2j-1$, $2j$, $2j-1+N$, and $2j+N$ are all occupied and where $\tau$ corresponds to the number of diagonal configurations in which either $2j-1/2j+N$ or $2j-1+N/2j$ are both occupied where $2j-1$ and $2j$ are adjacent, BCS-paired orbitals.  A few configurations with the necessary quantum numbers specified for $N=8$ are included in Fig. \ref{fig:N8arrows}.

As can be seen from Fig. \ref{fig:N8hist}, the groundstate wavefunction for the $N=8$ FEC Hamiltonian no longer simply contains elements of the BCS wavefunction and the Lipkin wavefunction naively entangled together.  Specifically, while the $|4,4,F,1,2\rangle$ class of basis states does include BCS-paired particles (see Fig. \ref{fig:N8arrows}), it does not include the maximal number of BCS-paired particles, which appears to be a necessary condition for non-zero occupation of the ground state for the BCS Hamiltonian.  However, this class of basis states can interact with other BCS-like and Lipkin-like classes of basis states.  Explicitly, $|4,4,F,1,2\rangle$ interacts with $|2,4,F\rangle$ via $\frac{\lambda}{2}\hat{a}^\dagger_p\hat{a}^\dagger_q\hat{a}_{q+N}\hat{a}_{p+N}$; $|4,4,F,1\rangle$ via $\frac{\lambda}{2}\hat{a}^\dagger_p\hat{a}^\dagger_{q+N}\hat{a}_{q}\hat{a}_{p+N}$; $|6,4,F\rangle$ via $\frac{\lambda}{2}\hat{a}^\dagger_{p+N}\hat{a}^\dagger_{q+N}\hat{a}_{q}\hat{a}_{p}$; and $|2,2,T\rangle$ via $-G\hat{a}^\dagger_{2j-1}\hat{a}^\dagger_{2j}\hat{a}_{2k}\hat{a}_{2k-1}$, which does further entangle the Lipkin-like configurations and BCS-like configurations in a non-trivial manner.  As such, while the interaction between the BCS-like classes of basis states and Lipkin-like classes of basis states in the formation of the FEC ground state wavefunction is not as clear-cut or simple as in the $N=4$ case, the $N=8$ FEC wavefunction is still an entanglement of BCS-like and Lipkin-like terms.

A representative configuration as well as the relevant quantum numbers for all classes of basis states for the $N=6$, $N=8$, and $N=12$ FEC Hamiltonians is given in the Supplemental Information.

\section{Discussion and Conclusions}
In this study, we introduce a model Hamiltonian that successfully demonstrates the physics associated with both fermion-pair condensation and exciton condensation, as well as encompassing the phase space consisting of systems in which fermion-pair condensation and exciton condensation are simultaneously realized---a phenomenon which we term fermion-exciton condensation (FEC). Applying this model to systems composed of $N=4,6,8,10$ particles in $r=2N$ orbitals, we confirm this fermion-exciton condensate character for a wide variety of ground state wavefunctions corresponding to a diverse range of input parameters in the model Hamiltonian, additionally verifying the prediction made in prior investigation \cite{Sager_2019} that the wavefunction of a fermion-exciton condensate is an entanglement of wavefunctions of exciton condensates and fermion-pair condensates.

The introduction of our model Hamiltonian that supports fermion-exciton condensation advances our understanding of the forces and orbital correlations necessary for the experimental construction of FEC states in real-world materials---important insights in the search for real-world materials exhibiting fermion-exciton condensate character.  Depending on the interpretation of the Hamiltonian elements, this could have ramifications for fields such as traditional and molecularly-scaled electronics, spin systems, and nuclear physics.

Specifically, if the orbitals in the Hamiltonian are interpreted as spin orbitals, fermion-exciton condensates simultaneously demonstrate the condensation of Cooper into a single particle-particle quantum state and the condensation of electron-hole pairs into a single particle-hole quantum state; thus, superfluid Cooper pairs---resulting in superconductivity---and superfluid excitons---which are associated with the dissipationless flow of energy \cite{keldysh_2017,Fil_Shevchenko_Rev}---should both be present to a certain extent in FEC systems, maybe demonstrating some hybridization of the properties of superconductors and exciton condensates, which may be relevant to the fields of energy transport and electronics in both macroscopic materials and molecular-scaled systems.

Alternatively, the two Lipkin-like $N$-degenerate levels can be interpreted as being representative of specific spin states such that the upper level is spin up and the lower level is spin down or vice versa.  This interpretation is most-consistent with $\epsilon=0$---which does demonstrate FEC states for a wide variety of input parameters---, although in a magnetic field the different spin states could be separated by some non-zero energy.  In this framework, the Lipkin-like terms could represent simultaneous double spin flips that are either aligned ($\lambda$) or misaligned ($\gamma$), and the pairwise Pairing-Force term could be seen as a favorable interaction between adjacent particles demonstrating the same spin.

Moreover, as both particle-particle (consistent with the Pairing-Force Hamiltonian) and particle-hole (consistent with the Lipkin Hamiltonian) are utilized in the field of nuclear physics to display the essential properties of the nuclear interaction \cite{ring_schuck_2004,cambiaggio_1997,Isacker:2014}, we can interpret our FEC Hamiltonian in this framework.  In this interpretation, the particles being created and annihilated are nucleons such that the Lipkin terms are associated with the interaction of nucleons within a valence shell ($\gamma$), the mixing of particle-hole excitations with the valence configurations, and excitations of a nucleon from one valence shell to another having an energetic penalty ($\epsilon$) \cite{ring_schuck_2004,Isacker:2014}.  Additionally, in this interpretation, the PF pairwise interaction is associated with the short-range portion of the nuclear interaction \cite{ring_schuck_2004,cambiaggio_1997}.

Overall, this model Hamiltonian is capable of demonstrating a wider array of collective behavior than either the Lipkin or the Pairing-Force models.  Such a Hamiltonian will have a vast degree of applications and will be beneficial for the exploration---and for benchmarking computational methodologies for the treatment of---the nontrivial physics of real-world material and chemical systems.

\vspace{0.25cm}
\noindent {\bf Author contributions.} L. M. and D. M. conceived of the project, developed the theoretical framework, designed the computations, wrote the code, performed the computations, analyzed the results, and wrote the paper.

\noindent
\textbf{Acknowledgments}: D.A.M. gratefully acknowledges the U.S. National Science Foundation Grants No. CHE-1565638, No. CHE-2035876, and No. DMR-2037783 and the Department of Energy, Office of Basic Energy Sciences, Grant DE-SC0019215.

\noindent
{\bf Data availability.} Data will be made available upon reasonable request.

\noindent
{\bf Code availability.} Code will be made available on a public Github repository upon publication.

\begin{appendix}
\section{Determination of Signatures of Condensation}
\label{app:signatures}
To determine the largest eigenvalue of the particle-particle RDM (${}^{2}D$, see Eq. (\ref{eq:D2})---i.e., $\lambda_D$, the signature of superconducting character---, only the following $N\times N$ subblock of the full 2-RDM containing the large eigenvalue must be computed and diagonalized \cite{Coleman_1965,Kade_2017,Poelmans_2015}
\begin{widetext}
\begin{equation}
	\begin{array}{c|cccc}
                                             &  { \hat{a}_{0}\hat{a}_{1}} &  { \hat{a}_{2}\hat{a}_{3}} & {\cdots}  &  { \hat{a}_{r-2}\hat{a}_{r-1}}\\ \hline
   { \hat{a}_{0}^\dagger\hat{a}^\dagger_{1}} &{ \hat{a}_{0}^\dagger\hat{a}^\dagger_{1}\hat{a}_{0}\hat{a}_{1}}  &{ \hat{a}_{0}^\dagger\hat{a}^\dagger_{1}\hat{a}_{2}\hat{a}_{3}}  &{\cdots}  &{ \hat{a}_{0}^\dagger\hat{a}^\dagger_{1}\hat{a}_{r-2}\hat{a}_{r-1}}  \\
   { \hat{a}_{2}^\dagger\hat{a}^\dagger_{3}} &{ \hat{a}_{2}^\dagger\hat{a}^\dagger_{3}\hat{a}_{0}\hat{a}_{1}}  &{ \hat{a}_{2}^\dagger\hat{a}^\dagger_{3}\hat{a}_{2}\hat{a}_{3}}  &{\cdots}  &{ \hat{a}_{2}^\dagger\hat{a}^\dagger_{3}\hat{a}_{r-2}\hat{a}_{r-1}}  \\
   {\vdots}                                  &{\vdots}  &{\vdots}  &{\ddots}  &{\vdots}  \\
   { \hat{a}_{r-2}^\dagger\hat{a}^\dagger_{r-1}} &{ \hat{a}_{r-2}^\dagger\hat{a}^\dagger_{r-1}\hat{a}_{0}\hat{a}_{1}}  &{ \hat{a}_{r-2}^\dagger\hat{a}^\dagger_{r-1}\hat{a}_{2}\hat{a}_{3}}  &{\cdots}  &{ \hat{a}_{r-2}^\dagger\hat{a}^\dagger_{r-1}\hat{a}_{r-2}\hat{a}_{r-1}}  \\
\end{array}
\label{eq:D2_submat}
\end{equation}
\end{widetext}
where, again, $\hat{a}^\dagger_i$ and $\hat{a}_i$ are to creation and annihilation operators corresponding to the orbital with index $i$.  Each element of this subblock of the 2-RDM is the expectation value $\langle\Psi|\hat{a}^\dagger_{2j-1}\hat{a}^\dagger_{2j}\hat{a}_{2k}\hat{a}_{2k-1}|\Psi\rangle$ obtained by programmatically applying the appropriate creation and annihilation operators to each pair of non-zero basis states composing the previously-obtained ground state wavefunction of the Hamiltonian.  As an example, for the $N,r=4$ computations, there are ten non-zero basis elements composing five distinct classes ($|0,2,T\rangle,|2,2,F\rangle,|2,2,T\rangle,|2,0,T\rangle,|4,2,T\rangle$) that are used to construct the Hamiltonian (see the Result section).  The ground-state wavefunction is obtained in terms of these classes with a structure given by
\begin{gather}
|\Psi\rangle=v_{0,2,T}|0,2,T\rangle+v_{2,2,F}|2,2,F\rangle+v_{2,2,T}|2,2,T\rangle\nonumber\\+v_{2,0,T}|2,0,T\rangle+v_{4,2,T}|4,2,T\rangle
\end{gather}
where each of the classes is a weighted linear combination of the basis states composing it, i.e,
\begin{equation}
|2,0,T\rangle=\frac{|1,3,6,8\rangle+|1,4,6,7\rangle+|2,3,5,8\rangle+|2,4,5,7\rangle}{\sqrt{4}}
\end{equation}
Thus, $\langle\Psi|\hat{a}^\dagger_{2j-1}\hat{a}^\dagger_{2j}\hat{a}_{2k}\hat{a}_{2k-1}|\Psi\rangle$ is a sum of all expectation values of the form
\begin{equation}
v_{c_1}v_{c_2}\langle c_1|\hat{a}^\dagger_{2j-1}\hat{a}^\dagger_{2j}\hat{a}_{2k}\hat{a}_{2k-1}|c_2\rangle
\end{equation}
where $c_1$ and $c_2$ refer to each of the distinct classes of non-zero basis states and where these expectation values are sums over
\begin{equation}
\frac{v_{b_1}v_{b_2}}{N(c_{b_1})N(c_{b_2})}\langle b_1|\hat{a}^\dagger_{2j-1}\hat{a}^\dagger_{2j}\hat{a}_{2k}\hat{a}_{2k-1}|b_2\rangle
\end{equation}
where $b_1$ and $b_2$ are the basis states composing each class, where $N(c_{b_1})$ refers to the size of the class to which basis $b_1$ belongs, and where all possible combinations of basis states are analyzed.

Note that only $\epsilon=0$ calculations were run for the $N,r=10,20$ scan such that site symmetry allowed the entire matrix to be constructed from three distinct types of elements, which lowered computational expense; these element types are as follows: $\hat{a}^\dagger_{2j-1}\hat{a}^\dagger_{2j}\hat{a}_{2j}\hat{a}_{2j-1}$, $\hat{a}^\dagger_{2j-1}\hat{a}^\dagger_{2j}\hat{a}_{2k}\hat{a}_{2k-1}$, and $\hat{a}^\dagger_{2j-1}\hat{a}^\dagger_{2j}\hat{a}_{2j\pm N}\hat{a}_{2j-1 \pm N}$.

The signature of superconductivity ($\lambda_D$) is then computed from the $N\times N$ subblock of the 2-RDM according to the eigenvalue equation
\begin{equation}
    {}^{2}Dv_D^i=\epsilon_D^iv_D^i
\end{equation}
with the signature corresponding the largest eigenvalue (the maximum $\epsilon_D^i$).

The portion of the particle-hole RDM (${}^{2}G$) associated with a large eigenvalue is composed of sub-matrices of the form
\begin{widetext}
\begin{equation}
\begin{array}{c|cccc}
     &  { \hat{a}_{q}^\dagger\hat{a}_{q}} &  { \hat{a}_{q+N}^\dagger\hat{a}_{q}} &  { \hat{a}_{q}^\dagger\hat{a}_{q+N}} &  { \hat{a}_{q+N}^\dagger\hat{a}_{q+N}}\\\hline
   { \hat{a}_{p}^\dagger\hat{a}_{p}} & {  \hat{a}_{p}^\dagger\hat{a}_{p}\hat{a}_{q}^\dagger\hat{a}_{q}}  &  { \hat{a}_{p}^\dagger\hat{a}_{p}\hat{a}_{q+N}^\dagger\hat{a}_{q}} &  { \hat{a}_{p}^\dagger\hat{a}_{p}\hat{a}_{q}^\dagger\hat{a}_{q+N}} &  { \hat{a}_{p}^\dagger\hat{a}_{p}\hat{a}_{q+N}^\dagger\hat{a}_{q+N}}\\
  {  \hat{a}_{p}^\dagger\hat{a}_{p+N}} &  { \hat{a}_{p}^\dagger\hat{a}_{p+N}\hat{a}_{q}^\dagger\hat{a}_{q}}  &  { \hat{a}_{p}^\dagger\hat{a}_{p+N}\hat{a}_{q+N}^\dagger\hat{a}_{q}} &  { \hat{a}_{p}^\dagger\hat{a}_{p+N}\hat{a}_{q}^\dagger\hat{a}_{q+N}} &  { \hat{a}_{p}^\dagger\hat{a}_{p+N}\hat{a}_{q+N}^\dagger\hat{a}_{q+N}}\\
   { \hat{a}_{p+N}^\dagger\hat{a}_{p}} &  { \hat{a}_{p+N}^\dagger\hat{a}_{p}\hat{a}_{q}^\dagger\hat{a}_{q}}  &  { \hat{a}_{p+N}^\dagger\hat{a}_{p}\hat{a}_{q+N}^\dagger\hat{a}_{q}} &  { \hat{a}_{p+N}^\dagger\hat{a}_{p}\hat{a}_{q}^\dagger\hat{a}_{q+N} }& {  \hat{a}_{p+N}^\dagger\hat{a}_{p}\hat{a}_{q+N}^\dagger\hat{a}_{q+N}}\\
   { \hat{a}_{p+N}^\dagger\hat{a}_{p+N}} &  { \hat{a}_{p+N}^\dagger\hat{a}_{p+N}\hat{a}_{q}^\dagger\hat{a}_{q}}   &  { \hat{a}_{p+N}^\dagger\hat{a}_{p+N}\hat{a}_{q+N}^\dagger\hat{a}_{q}} &  { \hat{a}_{p+N}^\dagger\hat{a}_{p+N}\hat{a}_{q}^\dagger\hat{a}_{q+N}} &  { \hat{a}_{p+N}^\dagger\hat{a}_{p+N}\hat{a}_{q+N}^\dagger\hat{a}_{q+N}}.
\end{array}
\end{equation}
tiled in the following manner:
\begin{equation}
\begin{array}{|c|c|c|c|}
	\hline
	{\scriptstyle  p=0,q=0} & {\scriptstyle  p=0, q=1} & \cdots & {\scriptstyle  p=0,q=\frac{N}{2}-1} \\\hline
	{ \scriptstyle p=1,q=0} & { \scriptstyle p=1, q=1} &  \cdots & { \scriptstyle p=1,q=\frac{N}{2}-1} \\\hline
	\vdots & \vdots & \ddots & \vdots \\\hline
	{ \scriptstyle p=\frac{N}{2}-1,q=0 }& { \scriptstyle p=\frac{N}{2}-1, q=1} & \cdots &{ \scriptstyle  p=\frac{N}{2}-1,q=\frac{N}{2}-1} \\\hline
	\end{array}
\end{equation}
\end{widetext}
In order to remove the ground-state-to-ground-state transition (to form the modified particle-hole RDM, ${}^{2}\Tilde{G}$, see Eq. (\ref{eq:modG2})),
\begin{widetext}
\begin{equation*}
\begin{array}{c|cccc}
     &  {\hat{a}_{q}^\dagger\hat{a}_{q}} &  {\hat{a}_{q+N}^\dagger\hat{a}_{q}} &  {\hat{a}_{q}^\dagger\hat{a}_{q+N}} &  {\hat{a}_{p+N}^\dagger\hat{a}_{p+N}}\\\hline
   { \hat{a}_{p}^\dagger\hat{a}_{p}} &  { {}^{1}D_p[0,0]{}^{1}D_q[0,0]} &   { {}^{1}D_p[0,0]{}^{1}D_q[0,1]} &   { {}^{1}D_p[0,0]{}^{1}D_q[1,0]} &  { {}^{1}D_p[0,0]{}^{1}D_q[1,1]} \\
  {  \hat{a}_{p}^\dagger\hat{a}_{p+N}} &  { {}^{1}D_p[0,1]{}^{1}D_q[0,0]} &   { {}^{1}D_p[0,1]{}^{1}D_q[0,1]} &   { {}^{1}D_p[0,1]{}^{1}D_q[1,0]} &  { {}^{1}D_p[0,1]{}^{1}D_q[1,1]}  \\
   { \hat{a}_{p+N}^\dagger\hat{a}_{p}} &  { {}^{1}D_p[1,0]{}^{1}D_q[0,0]} &   { {}^{1}D_p[1,0]{}^{1}D_q[0,1]} &   { {}^{1}D_p[1,0]{}^{1}D_q[1,0]} &  { {}^{1}D_p[1,0]{}^{1}D_q[1,1]} \\
   { \hat{a}_{p+N}^\dagger\hat{a}_{p+N}} &  { {}^{1}D_p[1,1]{}^{1}D_q[0,0]} &   { {}^{1}D_p[1,1]{}^{1}D_q[0,1]} &   { {}^{1}D_p[1,1]{}^{1}D_q[1,0]} &  { {}^{1}D_p[1,1]{}^{1}D_q[1,1]}
\end{array}
\end{equation*}
\end{widetext}
is subtracted off from each segment defined by $p$ and $q$ where the one-particle density matrix (${}^{1}D$) is given by
\begin{equation}
    \begin{array}{c|cc}
         & \hat{a}_p & \hat{a}_{p+N}\\\hline
     \hat{a}^\dagger_p    &\hat{a}^\dagger_p\hat{a}_p  & \hat{a}^\dagger_p \hat{a}_{p+N} \\
     \hat{a}^\dagger_{p+N}    &\hat{a}^\dagger_{p+N}\hat{a}_p & \hat{a}^\dagger_{p+N}\hat{a}_{p+N}
    \end{array}
\end{equation}

The signature of exciton condensation ($\lambda_G$) is then obtained from the eigenvalue equation
\begin{equation}
{}^{2}\Tilde{G}v_G^i=\epsilon_G^iv_G^i
\end{equation}
with the signature corresponding the largest eigenvalue (the maximum $\epsilon_G^i$).

Again, for the $N,r=10,20, \ \epsilon=0$ calculations, site symmetry was utilized to decrease computational expense.  Only sub-matrices corresponding to diagonal sub-matrices $p=q$, sub-matrices for BCS-paired orbitals $p=2j-1, \ q=2j$, and for unpaired orbitals $p=2j-1, \ q\ne p \ne 2j$ needed to be computed.

\section{Plastino's Model}
\label{app:plastino}
In literature that dates back to the 1960s and continues to this day, Plastino and coworkers \cite{Plastino_1978,Plastino_2018,Plastino_2021} explore a model Hamiltonian that adds a pairing-force term to the Lipkin model in the context of nuclear physics.  Introducing the Plastino pairing-force term to the Lipkin Hamiltonian from Eq. (\ref{eq:LipkinHam})---which allows for slightly more flexibility than the formulation given in the Plastino literature as that literature is concerned only with the double excitation/de-excitation ($\lambda$) term and omits the scattering term ($\gamma$)---yields the following model Hamiltonian:
\begin{gather}
    \mathcal{H}_{P}=-\frac{\epsilon}{2}\sum\limits_{i=1}^N\hat{a}^\dagger_i\hat{a}_i+\frac{\epsilon}{2}\sum\limits_{i=1}^N\hat{a}^\dagger_{i+N}\hat{a}_{i+N}\nonumber\\+\frac{\lambda}{2}\sum\limits_{p=1}^N\sum\limits_{q=1}^N\hat{a}^\dagger_p\hat{a}^\dagger_{q}\hat{a}_{q+N}\hat{a}_{p+N}+\frac{\lambda}{2}\sum\limits_{p=1}^N\sum\limits_{q=1}^N\hat{a}^\dagger_{p+N}\hat{a}^\dagger_{q+N}\hat{a}_{q}\hat{a}_{p}\nonumber\\+\frac{\gamma}{2}\sum\limits_{p=1}^N\sum\limits_{q=1}^N\hat{a}^\dagger_{p+N}\hat{a}^\dagger_{q}\hat{a}_{q+N}\hat{a}_{p}+\frac{\gamma}{2}\sum\limits_{p=1}^N\sum\limits_{q=1}^N\hat{a}^\dagger_{p}\hat{a}^\dagger_{q+N}\hat{a}_{q}\hat{a}_{p+N}\nonumber\\-G\sum\limits_{p=1}^N\sum\limits_{q=1}^N\hat{a}^\dagger_{p+N}\hat{a}^\dagger_{p}\hat{a}_{q}\hat{a}_{q+N}
\label{eq:Plastino}
\end{gather}
While the form of this Hamiltonian is similar to the one we introduce in Eq. (\ref{eq:Model}), the difference is the orbitals which the pairing-force term ($G$) causes to be correlated in Cooper-like pairs.  Specifically, while our model Hamiltonian pairs adjacent qubits (see Fig. \ref{fig:model}), the Plastino Hamiltonian pairs orbitals with on the same Lipkin-like cite in different layers (i.e., stacked orbitals $p$ and $p+N$).

\begin{figure}[tbh!]
  \includegraphics[width=8.5cm]{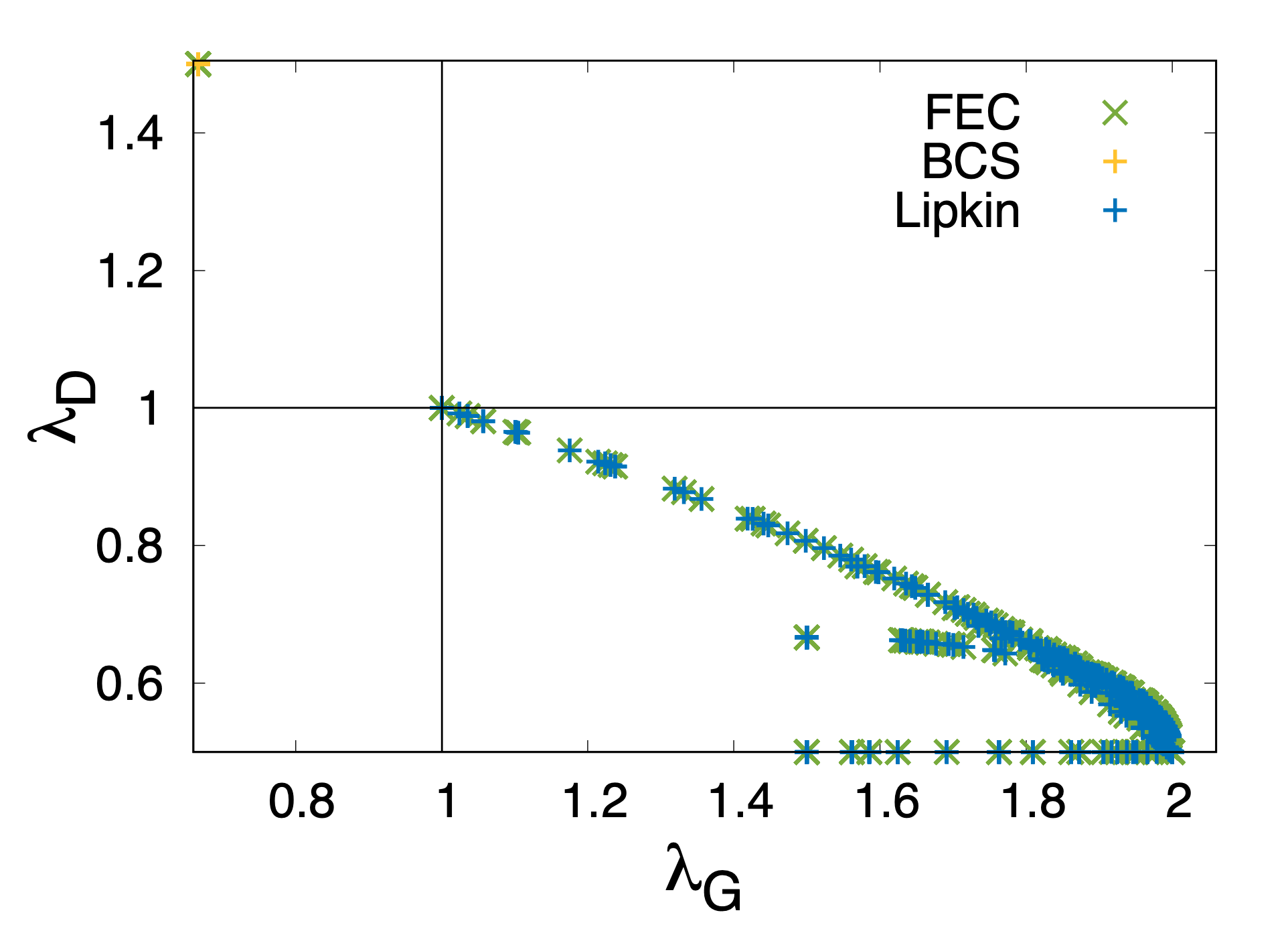}
  \caption{\label{fig:P_N4} A plot of $\lambda_G$ versus $\lambda_D$ where parameters in the Plastino Hamiltonian are systematically varied for $N=4$ particles in $r=8$ orbitals is shown.}
\end{figure}

In order to determine whether the Plastino Hamiltonian is capable of probing fermion-exciton condensate character---where $\lambda_D$ and $\lambda_G$ simultaneously exceed the Pauli-like limit of one and hence character of both fermion-pair condensation and exciton condensation are observed in a single quantum state---, a systematic scan over the input parameters of the Hamiltonian ($\epsilon,\lambda,\gamma,G$) is conducted.  As can in seen by Fig. \ref{fig:P_N4} where the blue pluses represent the Lipkin model Hamiltonian, the yellow pluses represent the PF BCS-like Hamiltonian, and the green x's represent the Plastino Hamiltonian, while Plastino's Hamiltonian is capable of reproducing all Lipkin states accessible by the Lipkin model and states that demonstrate fermion-pair condensation, no dual condensate character is observed from the Plastino model as the region in which both $\lambda_D$ and $\lambda_G$ exceed one is not probed within this model.

In fact, as noted in Ref. \onlinecite{}, there is direct competition between the particle-hole and particle-particle pairing between Lipkin-like sites which results in each type of pairing ``driving'' the system toward radically different states with the magnitudes of the coupling constants causing a transition between the Lipkin-like and BCS-like states favored by the different interactions.  Conversely, because the particle-particle and particle-hole pairing in the model we introduce do not occur between the same orbitals, they can coexist, allowing for a much larger possible range of $\lambda_D$ versus $\lambda_G$ including the region demonstrating a fermion-exciton condensate.

\end{appendix}

\bibliography{references,references_2}

\end{document}